\documentclass[aps,twocolumn,showpacs]{revtex4}
\usepackage{amsmath}
\usepackage{graphicx}
\usepackage{dcolumn}
\usepackage{verbatim}
\usepackage{times}
\usepackage{subfigure}
\usepackage{bm}
\usepackage{color}
\usepackage[colorlinks,dvipdfm]{hyperref}
\usepackage{subfigure}

\begin{document}

\title{Lifshitz transitions in a heavy-Fermion liquid driven by short-range
antiferromagnetic correlations in the two-dimensional Kondo lattice model}
\author{Guang-Ming Zhang$^1$, Yue-Hua Su$^2$, and Lu Yu$^3$}
\affiliation{$^1$Department of Physics, Tsinghua University, Beijing, 100084, China \\
$^{2}$Department of Physics, Yantai University, Yantai 264005, China\\
$^{3}$Institute of Physics and Institute of Theoretical Physics, Chinese
Academy of Sciences, Beijing 100190, China}
\date{\today}

\begin{abstract}
{The heavy-Fermion liquid with short-range antiferromagnetic correlations is
carefully considered in the two-dimensional Kondo-Heisenberg lattice model.
As the ratio of the local Heisenberg superexchange $J_{H}$ to the Kondo
coupling $J_{K}$ increases, Lifshitz transitions are anticipated, where the
topology of the Fermi surface (FS) of the heavy quasiparticles changes from
a hole-like circle to four kidney-like pockets centered around $(\pi ,\pi )$%
. In-between these two limiting cases, a first-order quantum phase
transition is identified at $J_{H}/J_{K}=0.1055$ where a small circle begins
to emerge within the large deformed circle. When $J_{H}/J_{K}=0.1425$, the
two deformed circles intersect each other and then decompose into four
kidney-like Fermi pockets via a second-order quantum phase transition. As $%
J_{H}/J_{K}$ increases further, the Fermi pockets are shifted along the
direction (}$\pi ,\pi ${) to (}$\pi /2,\pi /2${), and the resulting FS is
consistent with the FS obtained recently using the quantum Monte Carlo
cluster approach to the Kondo lattice system in the presence of the
antiferrmagnetic order.}
\end{abstract}

\pacs{64.70.Tg, 71.27.+a}
\maketitle

Quantum phase transitions are emergent phenomena observed in many strongly
correlated electron systems and has attracted much interest. An electronic
transition associated with the change of Fermi surface (FS) topology, the
so-called Lifshitz transition\cite{Lifshitz-1960}, can be induced without
any spontaneous symmetry breaking and local order parameter. The Lifshitz
transition is assumed to be a quantum phase transition at $T=0$ and it
becomes a crossover at finite temperatures. Elucidating the nature of the
Lifshitz transition, it seems that the transition manifests itself
dramatically\cite{Sandeman-2003,Yamaji-2007} only when other degrees of
freedom like lattice or spin couple strongly with the electronic states.

Heavy fermion materials have played a particularly important role in the
study of quantum critical phenomena. The Kondo lattice model is believed to
capture the basic physics of heavy fermions. The model describes a lattice
of local spin-1/2 magnetic moments coupled antiferromagnetically to a single
band of conduction electrons. The huge mass enhancement of the
quasiparticles can be attributed to the coherent superposition of individual
Kondo screening clouds, and the resulting metallic state is characterized by
a \textit{large} FS with the Luttinger volume containing both conduction
electrons and localized moments. Competing with the Kondo singlet formation,
the localized spins indirectly interact with each other via magnetic
polarization of the conduction electrons -- the
Ruderman-Kittel-Kasuya-Yosida interaction. Such interaction dominates at low
values of the Kondo exchange coupling and is the driving force for the
antiferromagnetic (AFM) long-range order quantum phase transitions\cite%
{Doniach,zhang-2000}. In a recent experiment\cite{Paschen-2004} a jump in
the Hall coefficient for YbRh$_{2}$Si$_{2}$ has been observed, and a sudden
change in the FS topology from a large FS to a small one was suggested at
the magnetic quantum critical point. The nature of this phase transition is
currently under hot debate\cite%
{Si-2001,Senthil-2004,Coleman-2005,Ogata-2007,Assaad-2008,
Pepin-2008,Senthil-2009}.

So far most of investigations focus on the possible FS reconstruction around
the magnetic quantum critical point. However, we would like to point out
that the FS topology in the paramagnetic heavy-fermion liquid phase may also
be drastically changed by the short-range AFM spin correlations between the
localized spins, leading to the Lifshitz phase transitions. The nature of
such a quantum phase transition has not been thoroughly explored yet. It is
well-known that the large-$N$ fermionic approach can be used to treat the
Kondo lattice model in the Kondo singlet regime very efficiently, leading to
the paramagnetic heavy-Fermion liquid state\cite%
{read-newns,Millis-1986,Auerbach-1986}. To consider the effects of the
short-range AFM spin correlations, it is more straightforward to explicitly
introduce the Heisenberg AFM superexchange $J_{H}$ between the localized
spins to the Kondo lattice system\cite%
{Coleman-1989,Senthil-2004,Coqblin-2003,Coleman-2005,Pepin-2008}.

In this paper, we apply the large-$N$ fermionic approach to the
Kondo-Heisenberg model on a two-dimensional square lattice in the limit of $%
J_{K}>J_{H}$. By introducing uniform short-range AFM valence-bond and Kondo
screening parameters, a fermionic mean-field theory is carefully
re-examined, and such a mean-field theory becomes exact when the degeneracy
of the localized spins $N$ becomes infinite. Away from half-filling, at the
conduction electron density $n_{c}=0.9$, for example, as $x=J_{H}/J_{K}$
increases, we find that the topology of the FS of the heavy quasiparticles
changes from one hole-like circle to four kidney-like pockets around ($\pi
,\pi $). In-between these two distinct limits, we will identify a
first-order quantum phase transition at $x_{1c}=0.1055$, where a small
circle begin to emerge within the large deformed circle. Then the inner
circle gradually expands, deforming to a rotated squared circle. When $%
x_{2c}=0.1425$, the two deformed circles intersect each other and then
decompose into four kidney-like Fermi pockets, resulting in a second-order
quantum phase transition.

The model Hamiltonian of the Kondo-Heisenberg lattice model is given by:
\begin{equation}
H=\sum_{\mathbf{k},\sigma }\epsilon _{\mathbf{k}}c_{\mathbf{k}\sigma
}^{\dagger }c_{\mathbf{k}\sigma }+J_{K}\sum_{i}\mathbf{S}_{i}\cdot \mathbf{s}%
_{i}+J_{H}\sum_{\left\langle ij\right\rangle }\mathbf{S}_{i}\cdot \mathbf{S}%
_{j},  \label{eqn1}
\end{equation}%
where $c_{k\sigma }^{\dagger }$ creates a conduction electron on an extended
orbital with wave vector $\mathbf{k}$ and z-component of spin $\sigma
=\uparrow ,\downarrow $. The spin-1/2 operators of the local magnetic
moments have the fermionic representation $\mathbf{S}_{i}=\frac{1}{2}%
\sum_{\sigma \sigma ^{\prime }}f_{i\sigma }^{\dagger }\mathbf{\tau }_{\sigma
\sigma ^{\prime }}f_{i\sigma ^{\prime }}$ with a local constraint $%
\sum_{\sigma }f_{i\sigma }^{\dagger }f_{i\sigma }=1$, where $\mathbf{\tau }$
is the Pauli matrices. Following the large-$N$ fermionic approach,\cite%
{Coleman-1989,Senthil-2004} the Kondo spin exchange and Heisenberg
superexchange terms can be expressed up to a chemical potential shift as
\begin{eqnarray}
\mathbf{S}_{i}\cdot \mathbf{S}_{j} &=&-\frac{1}{2}\sum_{\sigma \sigma
^{\prime }}f_{i\sigma }^{\dagger }f_{j\sigma }f_{j\sigma ^{\prime
}}^{\dagger }f_{i\sigma ^{\prime }},  \notag \\
\mathbf{S}_{i}\cdot \mathbf{s}_{j} &=&-\frac{1}{2}\sum_{\sigma \sigma
^{\prime }}f_{i\sigma }^{\dagger }c_{j\sigma }c_{j\sigma ^{\prime
}}^{\dagger }f_{i\sigma ^{\prime }},  \notag
\end{eqnarray}%
then a uniform short-range AFM valence bond and Kondo screening order
parameters can be introduced as
\begin{equation}
\chi =-\sum_{\sigma }\left\langle f_{i\sigma }^{\dagger }f_{i+l\sigma
}\right\rangle ,\text{ }V=\sum_{\sigma }\left\langle c_{i\sigma }^{\dagger
}f_{i\sigma }\right\rangle .  \label{eqn2}
\end{equation}%
Generally speaking, apart from the uniform short-range AFM valence bond
state, there are other possible competing states, including various flux
phases or plaquette states. However, the stabilization of those states would
require the presence of the translational lattice symmetry breaking and/or
additional frustrating interactions. Since there is no evidence of
translational symmetry breaking in the paramagnetic heavy-fermion liquid
states, we will not consider those effects at this stage. Although most
heavy fermion systems are three dimensional, as far as the FS topology is
concerned, it is conceptually simpler in the model discussion to start with
a two-dimensional case of this Kondo-Heisenberg lattice model.

To avoid the incidental degeneracy of the conduction electron band on a
square lattice, we choose $\epsilon _{\mathbf{k}}=-2t\left( \cos k_{x}+\cos
k_{y}\right) +4t^{\prime }\cos k_{x}\cos k_{y}-\mu $, where $t$ and $%
t^{\prime }$ are the first and second nearest neighbor hoping matrix
elements, respectively, while $\mu $ is the chemical potential, which should
be determined self-consistently by the density of the conduction electrons $%
n_{c}$. Under the uniform mean-field approximation, the f-fermions/spinons
form a very narrow band with the dispersion $\chi _{\mathbf{k}}=J_{H}\chi
\left( \cos k_{x}+\cos k_{y}\right) +\lambda $, where $\lambda $ is the
Lagrangian multiplier to be used to impose the local constraint on average.

In general, there are two interesting mean-field phases. One is the uniform
short-range AFM ordered phase in the limit of $J_{H}>J_{K}$ where $V=0$ but $%
\chi \neq 0$. In this phase, the spinons represented by f-fermions are
unconfined and have a dispersion $\chi _{\mathbf{k}}$. Another phase is the
heavy electron phase with short-range AFM spin correlations in the limit of $%
J_{H}<J_{K}$, where both $V\neq 0$ and $\chi \neq 0$. Actually the former
limit has been extensively studied in many previous investigations,\cite%
{Coleman-1989,Senthil-2004,Senthil-2009} but we will focus on the latter
limit.

Thus the corresponding mean-field Hamiltonian reads
\begin{equation}
H=\sum_{\mathbf{k}\sigma }\left(
\begin{array}{cc}
c_{\mathbf{k}\sigma }^{\dagger } & f_{\mathbf{k}\sigma }^{\dagger }%
\end{array}%
\right) \left(
\begin{array}{cc}
\epsilon _{\mathbf{k}} & -\frac{1}{2}J_{K}V \\
-\frac{1}{2}J_{K}V & \chi _{\mathbf{k}}%
\end{array}%
\right) \left(
\begin{array}{c}
c_{\mathbf{k}\sigma } \\
f_{\mathbf{k}\sigma }%
\end{array}%
\right) +E_{0},  \notag
\end{equation}%
with $E_{0}=N\left( -\lambda +J_{H}\chi ^{2}+J_{K}V^{2}/2\right) $. The
quasiparticle excitation spectra can be easily obtained
\begin{equation}
\varepsilon _{\mathbf{k}}^{\left( \pm \right) }=\frac{1}{2}\left[ \left(
\epsilon _{\mathbf{k}}+\chi _{\mathbf{k}}\right) \pm W_{\mathbf{k}}\right] ,
\label{eqn3}
\end{equation}%
which implies that the conduction electron band $\epsilon _{\mathbf{k}}$ has
a finite hybridization with the spinon band $\chi _{\mathbf{k}}$. Here $W_{%
\mathbf{k}}=\sqrt{\left( \varepsilon _{\mathbf{k}}-\chi _{\mathbf{k}}\right)
^{2}+\left( J_{K}V\right) ^{2}}$. Accordingly, the ground-state energy
density can be evaluated as
\begin{equation}
\varepsilon _{g}=\frac{2}{N}\sum_{\mathbf{k,\pm }}\varepsilon _{\mathbf{k}%
}^{\left( \pm \right) }\theta \left( -\varepsilon _{\mathbf{k}}^{\left( \pm
\right) }\right) -\lambda +J_{H}\chi ^{2}+\frac{1}{2}J_{K}V^{2},
\label{eqn4}
\end{equation}%
where $\theta \left( -\varepsilon \right) $ is the theta function. Then the
self-consistent equations for the mean-field variables $\chi $, $V$, and $%
\lambda $ can be derived by minimizing the ground state energy $\frac{%
\partial \varepsilon _{g}}{\partial \chi }=0,\text{ }\frac{\partial
\varepsilon _{g}}{\partial V}=0,\text{ }\frac{\partial \varepsilon _{g}}{%
\partial \lambda }=0$, and the chemical potential $\mu $ should be deduced
from the relation $n_{c}=-\frac{\partial \varepsilon _{g}}{\partial \mu }$.
This leads to the following self-consistent equations at zero temperature,
\begin{eqnarray}
\chi &=&-\frac{1}{2N}\sum_{\mathbf{k},\pm }\theta \left( -\varepsilon _{%
\mathbf{k}}^{\left( \pm \right) }\right) \left[ 1\mp \frac{\left( \epsilon _{%
\mathbf{k}}-\chi _{\mathbf{k}}\right) }{W_{\mathbf{k}}}\right] \gamma _{%
\mathbf{k}},  \notag \\
1 &=&-\frac{1}{N}\sum_{\mathbf{k},\pm }\theta \left( -\varepsilon _{\mathbf{k%
}}^{\left( \pm \right) }\right) \frac{\pm J_{K}}{W_{\mathbf{k}}},  \notag \\
1 &=&\frac{1}{N}\sum_{\mathbf{k},\pm }\theta \left( -\varepsilon _{\mathbf{k}%
}^{\left( \pm \right) }\right) \left[ 1\mp \frac{\left( \epsilon _{\mathbf{k}%
}-\chi _{\mathbf{k}}\right) }{W_{\mathbf{k}}}\right] ,  \notag \\
n_{c} &=&\frac{1}{N}\sum_{\mathbf{k},\pm }\theta \left( -\varepsilon _{%
\mathbf{k}}^{\left( \pm \right) }\right) \left[ 1\pm \frac{\left( \epsilon _{%
\mathbf{k}}-\chi _{\mathbf{k}}\right) }{W_{\mathbf{k}}}\right] ,  \label{eq5}
\end{eqnarray}%
where $\gamma _{\mathbf{k}}=\cos k_{x}+\cos k_{y}$. In the following, we
will assume that $t^{\prime }/t=0.3$ and $n_{c}=0.9$, which is away from
half-filling and in the paramagnetic metallic phase.\cite{Assaad-2008}

\begin{figure}[tbp]
\includegraphics[angle=90,scale=0.32]{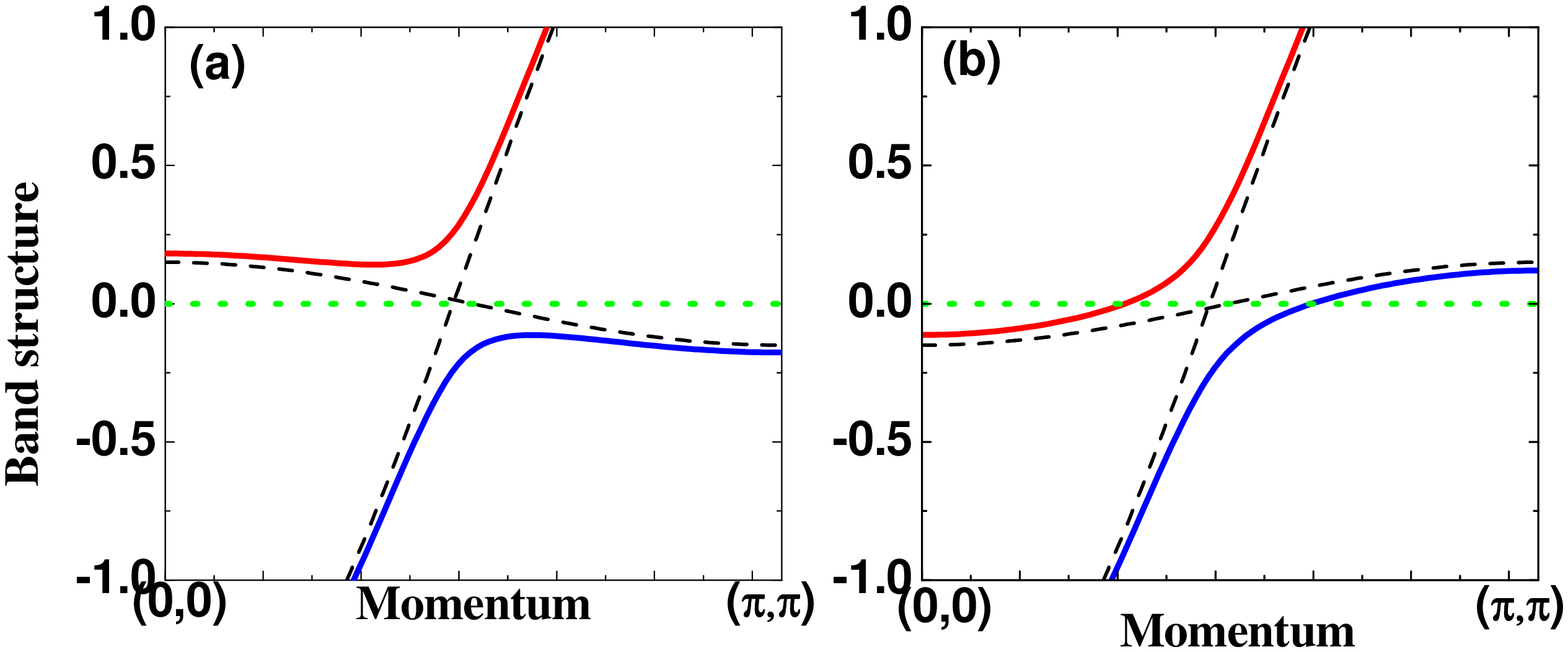}
\caption{(Color online) Schematic plot of the conduction electron band $%
\protect\epsilon _{\mathbf{k}}$ hybridizing with the very narrow band $%
\protect\chi _{\mathbf{k}}$ in the $n_{c}=1$ case. The dashed lines denote
the original bands, while the solid lines correspond to the hybridized
bands. (a) The short-range AFM order parameter $\protect\chi >0$, (b) $%
\protect\chi <0$. Clear distinction of these two different cases shows that
there is always a finite gap in case (a), while being gapless in case (b). }
\label{schematic-band}
\end{figure}

The most important thing is to notice that the resulting two renormalized
quasiparticle bands $\varepsilon _{\mathbf{k}}^{\left( \pm \right) }$
crucially depend on the sign of the AFM order parameter $\chi $. If $\chi $
is positive, $\varepsilon _{\mathbf{k}}^{\left( -\right) }$ and $\varepsilon
_{\mathbf{k}}^{\left( +\right) }$ are separated by an indirect energy gap as
displayed in Fig.1a, where the conduction electrons hybridize with the
\textit{hole-}like f-fermions/spinons. For the negative value of $\chi $,
however, the renormalized quasiparticle bands $\varepsilon _{\mathbf{k}%
}^{\left( -\right) }$ and $\varepsilon _{\mathbf{k}}^{\left( +\right) }$
always have a finite overlap as shown in Fig.1b, where the conduction
electrons actually hybridize with the \textit{particle-}like
f-fermions/spinons. Even in the half-filling case $n_{c}=1$, the model
system has the properties of a semi-metal with a FS consisting of one
electron-like pocket around $\mathbf{k}=(0,0)$ and one hole-like pocket
around $\mathbf{k}=(\pi ,\pi )$. Then the resulting ground state will
exhibit an instability towards the AFM spin-density wave or s$_{\pm }$-wave
paring superconductivity.\cite{chubukov-2008}

When the self-consistent calculations are carefully performed, we find that
the mean-field AFM order parameter $\chi $ is \textit{always} positive in
the range $0<J_{H}/J_{K}\leq 0.5$ so that the resulting state is a stable
paramagnetic metal. In Fig.2, as the strength of the AFM spin fluctuations
grows up, we present the evolution of the band structure of the renormalized
heavy quasiparticles around the Fermi level in the direction $(\pi /2,\pi
/2)\rightarrow (\pi ,\pi )\rightarrow (\pi ,\pi /2)$ of the first Brillouin
zone. Fig.2a to Fig.2i correspond to $J_{K}/t=2.0$ and $J_{H}/J_{K}=0.05$, $%
0.1055$, $0.1056$, $0.11$, $0.125$, $0.1375$, $0.1425$, $0.145$, and $0.25$,
respectively.

\begin{figure}[tbp]
\includegraphics[scale=0.43]{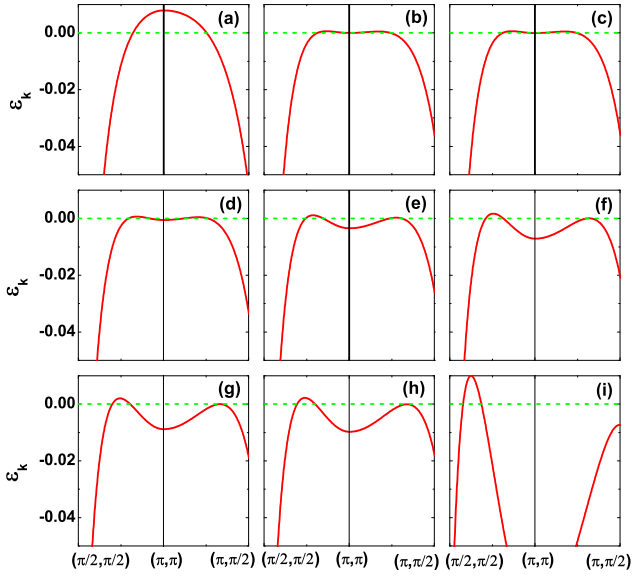}
\caption{(Color online) The lower renormalized quasiparticle band in the
direction $(\protect\pi /2,\protect\pi /2)\rightarrow (\protect\pi ,\protect%
\pi )\rightarrow (\protect\pi ,\protect\pi /2)$ as increasing the strength
of the AFM spin fluctuations. (a) to (i) correspond to $J_{K}/t=2.0$ and $%
J_{H}/J_{K}=0.05$, $0.1055$, $0.1056$, $0.11$, $0.125$, $0.1375$, $0.1425$, $%
0.145$, and $0.25$, respectively. In the calculation $t^{\prime }/t=0.3$ and
$n_{c}=0.9$. The dash line denotes the Fermi level.}
\label{Band-structure}
\end{figure}
For a fixed $J_{K}/t$, we can clearly see in Fig.2a and Fig.2b that the $M$
point $(\pi ,\pi )$ is the maximum of the lower renormalized quasiparticle
band $\varepsilon _{\mathbf{k}}^{\left( -\right) }$ for $J_{H}/J_{K}<0.1055$%
, and then the Fermi level ($\epsilon _{F}=0$) crosses the quasiparticle
band only once in the the direction $(\pi /2,\pi /2)\rightarrow (\pi ,\pi )$
or $(\pi ,\pi )\rightarrow (\pi ,\pi /2)$. For $J_{H}/J_{K}\geq 0.1055$, the
$M$ point becomes a local minimum of the lower renormalized quasiparticle
band $\varepsilon _{\mathbf{k}}^{\left( -\right) }$ in Fig.2c to Fig.2i. In
the direction of $(\pi /2,\pi /2)\rightarrow (\pi ,\pi )$, the Fermi level
always crosses the quasiparticle band twice. However, in the the direction
of $(\pi ,\pi )\rightarrow (\pi ,\pi /2)$, the Fermi level crosses the
quasiparticle band twice only when $0.1055\leq J_{H}/J_{K}<0.1425$, while
the Fermi level can cross the quasiparticle band once for $J_{H}/J_{K}\geq
0.1425$. Therefore, both $J_{H}/J_{K}=0.1055$ and $J_{H}/J_{K}=0.1425$
represent two special coupling strengths.

Once the renormalized quasiparticle band structure is available, the
corresponding FS can be easily obtained. Corresponding to the band structure
shown in Fig.2, the FS are displayed in Fig.3, where we have shifted the
center of the FS from ($0,0$) to ($\pi ,\pi $). For a fixed $J_{K}/t=2.0$,
we can clearly see that the FS is a hole-like circle around ($\pi ,\pi $)
for the parameter range $0\leq J_{H}/t\leq 0.1$, and then the shape of FS
deforms to a square for $0.1<J_{H}/J_{K}\leq 0.1055$. At $J_{H}/J_{K}=0.1056$%
, the topology of the FS starts to change: a small circle emerges in the
center of the deformed large square FS. As $J_{H}/J_{K}$ is further
increased, both circles expand and the small one is deformed into a rotated
square. Up to $J_{H}/J_{K}=0.1425$, the two deformed circles intersect each
other and then decompose into four kidney-like Fermi pockets. When $%
J_{H}/J_{K}$ continues to increase, the resulting FS (not included here)
will be shifted outward along the direction $M\longrightarrow \Gamma $.

Recently, a quantum Monte Carlo cluster approach has been proposed to study
the evolution of the Fermi surface across the magnetic order-disorder
transition in the two-dimensional Kondo lattice system\cite{Assaad-2008}. In
the AFM long-range ordered phase, the Kondo screening does not break down,
and the heavy fermion bands drop below the FS giving way to hole pockets
centered around $\mathbf{k}=(\pi /2,\pi /2)$ and equivalent points. These
results are fully consistent with the FS obtained by our calculation in the
range $0.25\leq J_{H}/J_{K}<1$.

\begin{figure}[tbp]
\begin{center}
\includegraphics[scale=0.4]{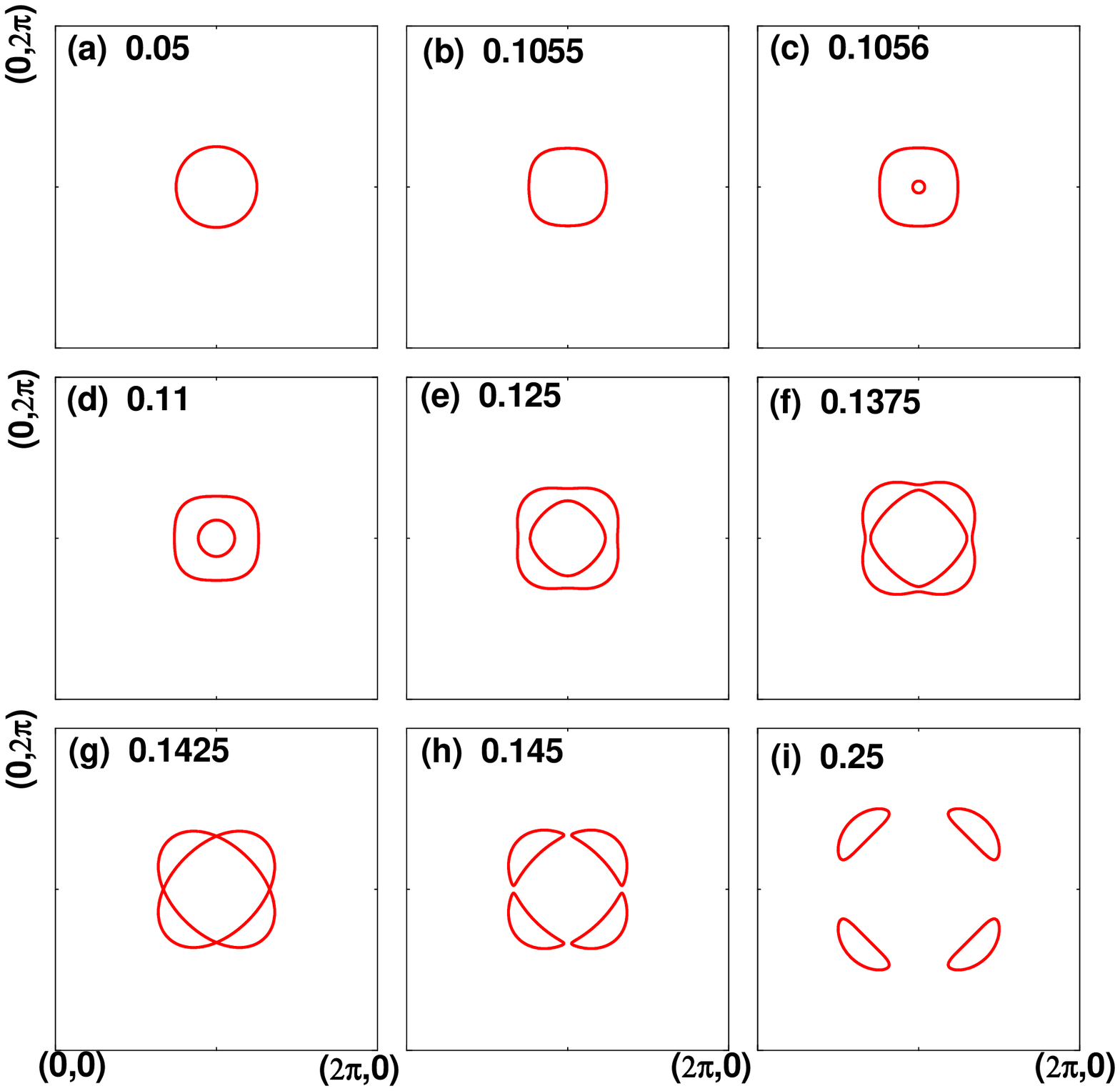}
\end{center}
\caption{(Color online) The Fermi surface corresponding to the lower
renormalized quasiparticle band shown in Fig.2. The ratio $J_{H}/J_{K}$ is
indicated in graphes. Here the center of the Fermi surface has been shifted
from $(0,0)$ to $(\protect\pi ,\protect\pi )$ in order to clearly exhibit
the topology of the Fermi surface.}
\label{Fermi-surface}
\end{figure}

In order to study the topology changes of the FS from the quantum phase
transition aspects, we calculate the ground state energy density $%
\varepsilon _{g}$ and its first-order derivative with respect to the ratio
of the coupling parameters $x=J_{H}/J_{K}$. The numerical results are
displayed in Fig.4a and Fig.4b. We find that two non-analytical points
appear in the ground state energy density. $\varepsilon _{g}$ is finite and
continuous in the parameter range $0<x<0.5$, However, its first-order
derivative has a large jump at $x_{1c}=0.1055$, corresponding to a
first-order (discontinuous) quantum phase transition. Moreover, a small kink
appears at $x_{2c}=0.1425$ in the first-order derivative, which corresponds
to a jump in the second-order derivative of $\varepsilon _{g}$. So $x_{2c}$
denotes a second-order (continuous) quantum phase transition. Actually both
quantum phase transitions belong to the category of Lifshitz phase
transitions. The ground state phase diagram is delineated in Fig.4c, where
there exist three different paramagnetic heavy-fermion liquid phases: the
conventional heavy-fermion liquid in $x<0.105$, the heavy-fermion liquid
with strong AFM spin fluctuations in $0.1425<x<1$, and the intermediate
phase $0.105<x<0.1425$. \textbf{\ }

\begin{figure}[tbp]
\begin{center}
\includegraphics[angle=90,scale=0.32]{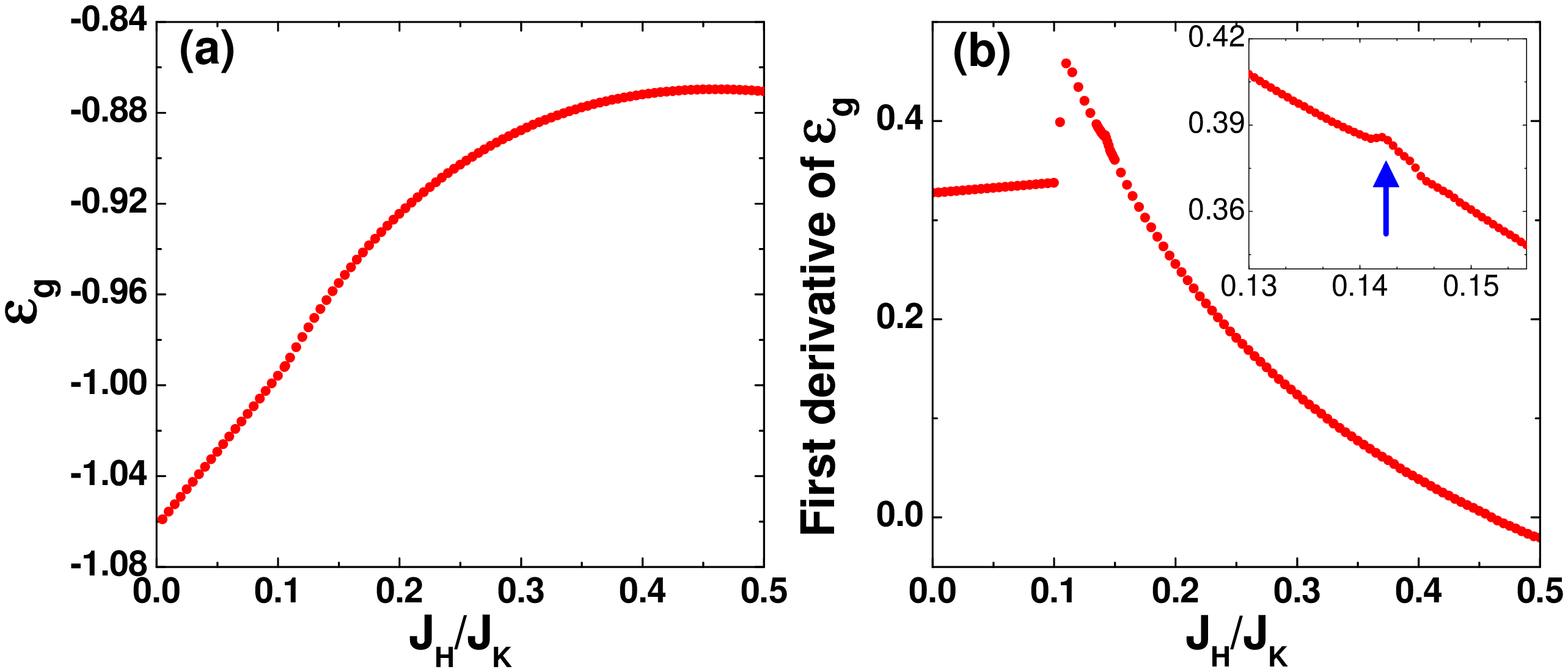}\newline
\includegraphics[angle=0,scale=0.32]{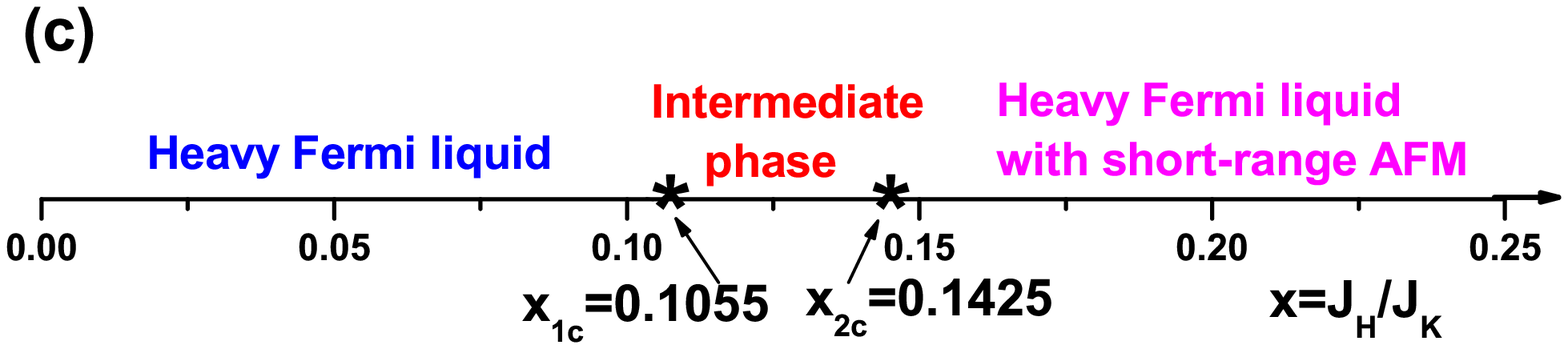}
\end{center}.
\caption{(Color online) (a) The ground state energy density $\protect%
\varepsilon _{g}$ as a function of $x=J_{H}/J_{K}$. (b) The first-order
derivative of $\protect\varepsilon _{g}$ with respect to the parameter $x$.
(c) The ground-state phase diagram of the system.}
\label{Ground-energy}
\end{figure}

\begin{figure}[tbp]
\begin{center}
\includegraphics[angle=90,scale=0.3]{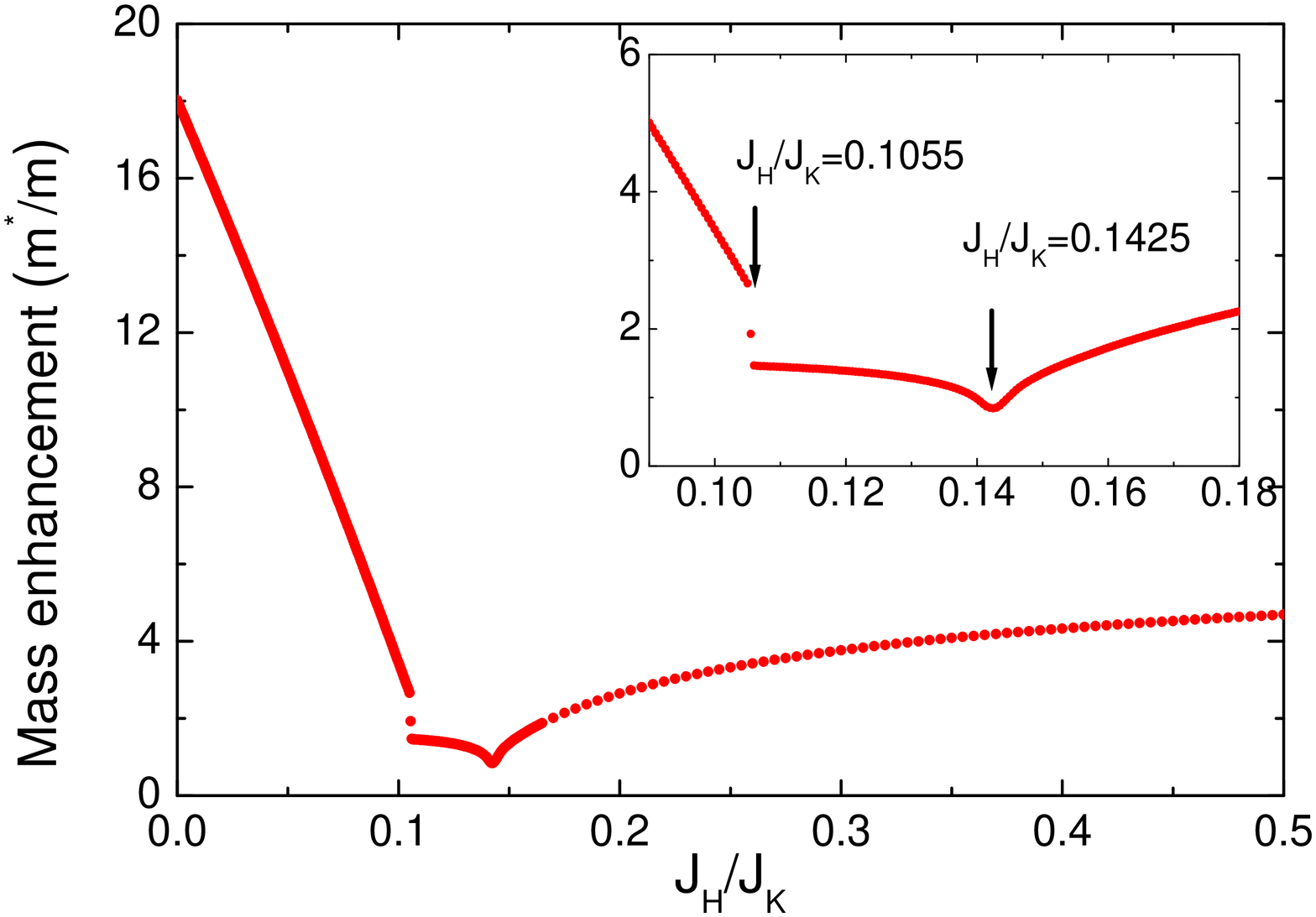}
\end{center}
\caption{(Color online) The quasiparticle mass enhancement factor
as a function of $x=J_{H}/J_{K}$. } \label{quasiparticle-mass}
\end{figure}

Moreover, the effective mass of the heavy quasiparticle excitations is a
function of the band curvature, so the topological changes of the Fermi
surface can be reflected in the effective mass, which is related to the
non-interacting band mass by the variation $m^{\ast }\delta \varepsilon _{%
\mathbf{k}}^{\left( -\right) }=m\delta \epsilon _{\mathbf{k}}$, averaged
over all points on the FS. Hence the mass enhancement factor is given by%
\begin{equation}
\frac{m^{\ast }}{m}=\left\langle \frac{\partial \varepsilon _{\mathbf{k}%
}^{\left( -\right) }}{\partial \epsilon _{\mathbf{k}}}\right\rangle
_{FS}^{-1}=\frac{1}{N}\sum_{\mathbf{k}}\left[ \frac{\partial \varepsilon _{%
\mathbf{k}}^{\left( -\right) }}{\partial \epsilon _{\mathbf{k}}}\right]
^{-1}\delta \left( \mu -\varepsilon _{\mathbf{k}}^{\left( -\right) }\right) ,
\label{eqn6}
\end{equation}%
which is displayed in Fig.5. Here we can also observe the two
successive quantum phase transitions at $J_{H}/J_{K}=0.1055$ and
$0.1425$, respectively, consistent with the results from the
analysis of the ground state energy density. Actually, since the
effective mass enhancement factor is related to the optical
conductivity in infrared spectroscopy measurements, the above
Lifshitz phase transitions can be observed experimentally.

In conclusion, we have carefully studied the heavy-fermion liquid state in
the two-dimensional Kondo-Heisenberg lattice system. As $J_{H}/J_{K}$ grows
up, the topology of the quasiparticle FS rapidly changes from one hole-like
circle in the conventional heavy-fermion liquid state to four kidney-like
pockets centered around $(\pi ,\pi )$, which is very close to the FS near
the AFM magnetic quantum critical point. Between these two distinct FSs, a
first-order quantum phase transition occurs at $J_{H}/J_{K}=0.1055$, where a
small circle emerges within the large deformed circle. When $%
J_{H}/J_{K}=0.1425$, the two deformed circles intersect each other and then
decompose into four kidney-like Fermi pockets, and a second-order quantum
phase transition takes place. Both quantum phase transitions belong to the
category Lifshitz phase transitions.

To some extent our present mean-field theory captures the heavy-fermion
liquid physics of the Kondo-Heisenberg lattice systems, especially the Fermi
surface evolution of the renormalized heavy quasiparticles as the
short-range AFM spin correlations between the localized magnetic moments are
gradually increased. In order to put the present results on a more solid
ground, further investigations including the gauge fluctuations associated
with the mean-field order parameters are certainly needed.

The authors would like to thank T. Xiang and D. H. Lee for their
stimulating discussions and acknowledge the support of NSF of
China and the National Program for Basic Research of MOST-China.

\end{document}